 \documentstyle[aps,eqsecnum,floats,epsf]{revtex}

\begin{document}

\draft

\title{Fate of Kaluza-Klein Bubble
\footnote{Preprint numbers: CGPG-99/12-8, RESCEU-6/00, 
DAMTP-2000-30, hep-th/0003066. \\
This article will appear in  Phys. Rev. D.}}
\author{Hisa-aki Shinkai  \cite{Email-his}}
\address{
Center for Gravitational Physics and Geometry,
104 Davey Lab., Department of Physics,\\
The Pennsylvania State University,
University Park, Pennsylvania 16802-6300
}

\author{Tetsuya Shiromizu \cite{Email-ts}}
\address{DAMTP, University of Cambridge,
Silver Street, Cambridge CB3 9EW, United Kingdom,  \\
Department of Physics, The University of Tokyo, Tokyo 113-0033,
Japan \\
and \\
Research Centre for the Early Universe(RESCEU),
The University of Tokyo, Tokyo 113-0033, Japan
}

\date{March 2, 2000 (revised version) 
}
\maketitle

\begin{abstract}

We numerically study classical time evolutions of Kaluza-Klein
bubble space-time which has negative energy after a decay of vacuum.
As the zero energy Witten's bubble space-time,
where the bubble expands infinitely,
the subsequent evolutions of Brill and Horowitz's momentarily static
initial data show that the bubble will expand in terms of the area.
At first glance, this result may support
Corley and Jacobson's conjecture that the
bubble will expand forever as well as the Witten's bubble.
The irregular signatures, however, can be seen in the behavior
of the lapse
function in the maximal slicing gauge
and the divergence of the Kretchman invariant.
Since there is no appearance of the apparent horizon, we
suspect an appearance of a naked singularity as the final fate of
this space-time.
\end{abstract}

\section{Introduction}

It is likely that superstring or M-theory governs the physics of
gravity or space-time in higher energy stages\cite{String}.
Such theories are naturally formulated in the higher dimensions than
four.  We expect a plausible scenario that such a higher dimensional
space-time somehow evolves to the stable four dimensional space-time
according to the history of the Universe
\cite{brane,Tess,recent}.
The so-called brane world scenario\cite{brane} is the most
actively being investigated along to this line.
This scenario is motivated by Horava and Witten's theory \cite{Horava}
which shows that
an eleven dimensional supergravity theory on the orbihold
${\bf R}^{10} \times S^1 /Z_2$ is related to the
ten-dimensional $E_8 \times E_8$
heterotic string theory. Therein the matters are confined to the
ten-dimensional
space-time (three-brane) and gravitons are propagating in
the full eleven
dimensions. The brane world space-time should be stable.

Although the brane world scenario may be plausible at
the reduction from eleven to ten dimensions, the space-time
will be still compactified to four dimensions in the
normal Calabi-Yau's way.
Regarding to these full scenario of the compactification,
the stability of the space-time becomes
the important issue to be investigated.
The positive energy theorem guarantees the stability of the
four dimensional
asymptotically flat space-time
in the framework of general relativity \cite{PET}.
Surprisingly, the existence of the extra dimensions can
drastically change the
situation. Witten showed that the five dimensional Minkowski
space-time decays into the so-called Kaluza-Klein (KK) bubble
space-time unless we assume the existence of the elementary fermion
related to supersymmetry \cite{Witten,Marika}, of which existence
we can not expect generally.
This also may indicate that the `bubble' appears
somewhere at the bulk or on the brane in the brane world scenario and
disturbed the three-brane where we are living.

The metric of the KK bubble space-time given by Witten
is written as\cite{Witten}
%
\begin{eqnarray}
ds_5^2 & = & -r^2dt^2+\Bigl(1-\frac{r^2_0}{r^2}\Bigr)d\chi^2+
\Bigl(1-\frac{r^2_0}{r^2}\Bigr)^{-1}dr^2 +r^2 \cosh^2 t \,
d\Omega^2,
\end{eqnarray}
%
where the $\chi$-direction will be compactified and $r \geq r_0$,
and $d\Omega^2=d\theta^2+\sin^2\theta d\varphi^2$.
In general case, the metric has a conical singularity at $r_0$.
However, if we carefully
take a periodicity along the $\chi$-direction, the metric
can be regularized.  More precisely to see this, we write
the metric near $r=r_0$ as
%
\begin{eqnarray}
ds_5^2 & \simeq & -r_0^2dt^2+\frac{2(r-r_0)}{r_0}d\chi^2+
\frac{r_0}{2(r-r_0)}dr^2 +r^2_0 \cosh^2 t \,
d\Omega^2
\nonumber \\
& = &  -r_0^2dt^2+2r_0\Bigl[R^2d\Bigl(\frac{\chi}{r_0}\Bigr)^2
+dR^2 \Bigr]
+r^2_0 \cosh^2 t \, d\Omega^2,
\end{eqnarray}
%
where $R=\sqrt{r-r_0}$.
Then we realize that the period should be set to
be $\chi_p=2\pi r_0$. As one can see, the `boundary of
bubble' located at $r=r_0$ expands rapidly like $\cosh t$ and
the space-time does not have naked singularities.
Here, we remind you that the total energy is zero.
We are imaging the boundary of the space-time, $r=r_0$, as the
surface of the `bubble'.

Interestingly Brill and Pfister \cite{Brill1} gave an initial
data which has the negative total energy related to the size
of the compactified dimension. The space-time with the negative energy
may be favor in the aspect of energetics.
One year later, Brill and Horowitz\cite{Brill2} gave
an initial data in a simple way.
(We will briefly review their construction in Sec. \ref{BH-initialdata}.)
Contrasted to the `Witten bubble', their solution has
arbitrary negative energy regardless of the size of the compactified
space.  This is too far from our intuition that the negative energy
is proportional to the Casimir energy due to the boundary effect of
the compactified space. Therefore it is difficult to imagine the
classical evolution after the vacuum decay.

Corley and Jacobson \cite{Ted} discussed the subsequent
evolution of Brill-Horowitz's initial data.
They found that the positive acceleration of the bubble's
surface area for
the negative mass bubbles, and they conjectured that
KK bubble with negative energy cannot collapse.
However, their study is not sufficient to conclude
the final fate of the bubble, as they already mentioned,
because they considered
only the initial behavior of the time-symmetric data and
they did not do any dynamical studies.

In this paper, we report our numerical analysis on this
final fate problem of KK bubble, especially of the negative
mass bubble.
We start our numerical simulation from
the Brill-Horowitz's initial data, and evolve the space-time
using the standard Arnowitt-Deser-Misner formulation (but 4+1
dimensional decomposition).
We will show that the space-time initially behaves as
Corley-Jacobson's analysis, and expands forever, although
the acceleration will be negative.
Despite of the expanding, we will
observe the irregular behavior of the curvature invariant.

This paper is organized as follows.
In Sec. \ref{BH-initialdata}, we
give a brief review of Brill-Horowitz's construction of their
initial data.
In Sec. \ref{Numerical-method}, we
describe numerical method and equations.
The results of our simulations are shown in
in Sec. \ref{Results}.
Finally, we summarize our results
 in Sec. \ref{Summary}.

\section{Brill-Horowitz's Initial Data} \label{BH-initialdata}

In this section we briefly review Brill and Horowitz's argument
\cite{Brill2}.
Let us consider an initial slice with $K_{ij}=0$ in five dimensional
vacuum space-times, where $K_{ij}$ is the
extrinsic curvature of a four dimensional
spacelike hypersurface. In this slice the Hamiltonian
constraint equation becomes
$ {}^{(4)}R=0$, where $ {}^{(4)}R$ is the four dimensional Ricci scalar.
 Here one can easily see that the Euclidean Reissner-Nordstrom
metric with imaginary `charge' $iq$ satisfies the Hamiltonian
constraint equation, because the `energy-momentum' tensor of the four
dimensional Maxwell field is traceless.
The metric of the hypersurface is given by
%
\begin{eqnarray}
{}^{(4)}g=U(r)d\chi^2+\frac{dr^2}{U(r)}+r^2
d\Omega^2, \label{eq:BH}
\end{eqnarray}
%
where $U(r)=1-2m/r-q^2/r^2$ and $r \geq r_+:=
m+{\sqrt {m^2+q^2}}$. In the same way as the previous Witten's
example, the metric is approximately written as
%
\begin{eqnarray}
{}^{(4)}g\simeq \frac{4}{U'(r_+)}\Bigl[R^2d
\Bigl(\frac{U'(r_+)\chi}{2} \Bigr)^2+dR^2  \Bigr]+r_+^2
d\Omega^2,
\end{eqnarray}
%
near $r=r_+$, where $R={\sqrt {r-r_+}}$.
To avoid a conical singularity at
$r=r_+$, we assume the period $\chi_p=4\pi/U'(r_+)=2\pi r_+^2/(r_+-m)$
along the $\chi$-direction.

The total energy is evaluated as $E=m/2$. The $m$ is arbitrary
parameter and $q$ determines the size of the compactified space.
So the total energy can be arbitrary negative.

\section{Field equations and our numerical method}
\label{Numerical-method}

To know the final fate of the KK bubbles with the
negative energy, we
study the subsequent time evolution for a long time numerically.
We apply 4+1 decomposition
of space-time along to the Arnowitt-Deser-Misner formulation
for the actual
time integrations.  We describe equations and basic numerical
techniques in this
section.

The metric of the full space-time is assumed to be
%
\begin{eqnarray}
ds_5^2& = & -N(r,t)^2dt^2+e^{2a(r,t)}U(r)d\chi^2
+e^{2b(r,t)}U(r)^{-1}dr^2+r^2e^{2c(r,t)}
d\Omega^2, \label{ourmetric}
\end{eqnarray}
%
where $U(r)=1-2m/r-q^2/r^2$,
$N$ is the lapse function, and the metric components $a,
b$ and $c$ are now time dependent.  The evolution equations of
the four-metric $\gamma_{ij}$ and the extrinsic
curvature $K_{ij}$ become
\footnote{Here,  for simplicity, we tacitly supposed the boundary condition
so that the location of the bubble is `fixed' under the variation of
the action. As a result we obtain the 5-dimensional vacuum
Einstein equation and can show the consistent result given in
Appendix A. Since the Cauchy development of the initial data
cannot cover all region outside the bubble, one may be able to
consider another boundary conditions, which might be artificial.}
%
\begin{eqnarray}
\dot{K}^i_j&=&N({}^{(4)}R^i_j+KK^i_j) -{}^{(4)}D^i{}^{(4)}D_jN,
\label{eq:einstein1}
\\
\dot{\gamma}_{ij}&=&-2NK_{ij}\label{eq:einstein2},
\end{eqnarray}
a dot denotes the time derivative, and ${}^{(4)}R^i_j$ and
${}^{(4)}D^i$ denote four dimensional Ricci curvature and
the covariant derivative,
respectively.
For the reader's convenience, we write down several terms in
(\ref{eq:einstein1})  for the metric (\ref{ourmetric}) as:
\begin{eqnarray}
{}^{(4)}R^\chi_\chi &= &e^{-2b} \Bigl[
\Bigl( -a'^2-a''-2a'c'-\frac{2a'}{r} +a'b'\Bigr)U
+\Bigl(-\frac{3}{2}a'+\frac{1}{2}b'-c'-\frac{1}{r}
 \Bigr)U'-\frac{1}{2}U''
\Bigr],\\
{}^{(4)}R^r_r& = & e^{-2b}\Bigl[ \Bigl(
-a'^2-a''-2c'^2-2c''-\frac{4c'}{r}
+a'b'+2b'c'+\frac{2b'}{r}
\Bigr)U
+\Bigl(-\frac{3}{2}a'+\frac{1}{2}b'-c'-\frac{1}{r}  \Bigr)
U'-\frac{1}{2}U''
\Bigr], \\
{}^{(4)}R^\theta_\theta & = & {}^{(4)}R^\varphi_\varphi
=e^{-2b}\Bigl[ \Bigl(-2c'^2-c''-\frac{4c'}{r}-\frac{1}{r^2}+c'b'
+\frac{b'}{r}
-a'c'-\frac{a'}{r} \Bigr)U
+\Bigl(-c'-\frac{1}{r} \Bigr)U'+\frac{e^{2b-2c}}{r^2}\Bigr],
\end{eqnarray}
and
\begin{eqnarray}
{}^{(4)}D_\chi {}^{(4)}D^\chi N & = &  
e^{-2b}\Bigl(a'+\frac{1}{2}\frac{U'}{U}  \Bigr)UN',\\
{}^{(4)}D_r {}^{(4)}D^r N & = &
e^{-2b}\Bigl[ \Bigl( N''-b'N'\Bigr)U +\frac{1}{2}N'U' \Bigr],\\
{}^{(4)}D_\theta {}^{(4)}D^\theta N & = &
e^{-2b}\Bigl(c'+\frac{1}{r} \Bigr)N'U,
\end{eqnarray}
where a dash denotes the derivative on $r$.

We start our simulation from the initial data of Brill-Horowitz's
momentarily static solution, such as
%
\begin{eqnarray}
a(r,0)=b(r,0)=c(r,0)&=&0, \label{eq:initial1} \\
K^\chi_\chi(r,0)=K^r_r(r,0)=K^\theta_\theta(r,0)&=&0.
\label{eq:initial2}
\end{eqnarray}
%

The numerical region is taken as $r_+ \leq r \leq r_e$, where
$r_+:= m+{\sqrt {m^2+q^2}}$ is the location of the bubble
at the initial data and $r_e$ is the numerical outer boundary.
We stress from the construction that the Kaluza-Klein bubble
space-time is restricted in $r_+ \leq r \leq \infty$\cite{Brill2}.
We apply the Robin boundary condition  at $r=r_e$ such as all the
components fall off as they form an
asymptotically flat spacetime.
At the inner boundary $r=r_+$, we use the fact that both
$a$ and $b$ evolve synchronously as we describe in the Appendix,
and use both the evolution equation for tr$K$,
\begin{equation}
\dot{K}=N \, K_{ij}K^{ij} -{}^{(4)}D^i{}^{(4)}D_i \, N,
\label{eq:trace}
\end{equation}
where we used the Hamiltonian constraint equation, and the momentum
constraint equation,
\begin{equation}
{}^{(4)}D_j K^j_r - {}^{(4)}D_r K =0,
\end{equation}
so as the system evolves properly.

In order to specify the lapse function, $N$,
we apply both the geodesic slicing condition, $N=1$,  and
the maximal slicing condition, $K=0$, which equation
becomes [directly from Eq. (\ref{eq:trace})]
\begin{equation}
{}^{(4)}\Delta N = N K_{ij}K^{ij}. \label{maximalslice}
\end{equation}
This elliptic equation is solved using the incomplete
Cholsky conjugate gradient method.  The outer boundary for $N$ is
set again as asymptotically flat, and the inner boundary at $r_+$
for solving (\ref{maximalslice}) we lineally extrapolate 4-metric
components.

We apply Brailovskaya integration scheme
(a second order predictor-corrector method) \cite{bernstein} for
the time evolution.
The accuracy of the calculation is checked by monitoring the
violation of the Hamiltonian constraint equation.  The numerical
code passed convergence tests, and
the results shown in this
paper are all obtained with acceptable accuracy.

\section{Results} \label{Results}
\subsection{Acceleration of the bubble surface}

We first check whether our code reveals the initial behavior
discussed by Corley and Jacobson \cite{Ted}.
We calculate 
 the area of the bubble, 
\begin{equation}
A(t)=4 \pi g_{\theta\theta}(r_+,t), \label{circumf}
\end{equation}
together with its time derivative $\dot{A}$, and its
acceleration,
\begin{equation}
\ddot{A}=4\pi \ddot{g}_{\theta\theta}= 4 \pi r^2 [
- 2 N \dot{K}^\theta_\theta - 2 \dot{N}{K}^\theta_\theta
+ 4 N^2 ({K}^\theta_\theta)^2 ] e^{2c}.  \label{accel}
\end{equation}
We first show this acceleration
in Fig.\ref{fig-acc}(a)(b), since this was the quantity discussed by
Corley and Jacobson.
The Figs. \ref{fig-acc}(a) and (b) are of the geodesic slicing
condition and
of the maximal slicing condition, respectively.
We fix the charge $q$ and varied $m$ from negative to positive
values. Except for the transition at $m=0$, as one can see later,
our result is not qualitatively sensitive under changes of $m/q$.
Under both slicing conditions,
we see that the negative mass bubble start expanding
(positive $\ddot{A}$) initially, yet
will soon be in de-accelerating
phase (negative $\ddot{A}$),
 while the positive mass bubble keep accelerating all the way in
Figs. \ref{fig-acc}(a), and in the region in Figs. \ref{fig-acc}(b).
More precisely, for the positive mass cases in Figs. \ref{fig-acc}(b),
we observe from the numerical results that 
the acceleration will reach and stay at a
positive value in the final stage, even if it goes negative for
a short time, which is happen to quite small positive mass cases.

Such an initial behavior (for both positive and negative mass bubbles)
does agree
with Corley and Jacobson's analysis (we remark that their analysis
was under the geodesic slicing condition).
However, the
turning behavior into de-accelarating phase could not find in
their analysis.  The de-accelarating does not mean collapsing feature
directly.  Actually, upto we stop our time evolution,
the numerical data of
the area, (\ref{circumf}),  monotonically increases [Figs.
\ref{fig-acc}(c)], while its velocity goes down for negative mass
bubbles [Figs. \ref{fig-acc}(d)].  However, from this facts, we can not say
that negative mass bubbles will expand forever, because we can see the
blow-ups of the  Riemann invariant and collapsing lapse behavior as we show
next.  (We had to stop time evolution for negative mass bubble case when
we face the blow-ups of the Riemann invariant.)

\subsection{Collapse of lapse}
Since we found that the time integration using the
maximal slicing condition survives long term time evolution than that
of the geodesic slicing condition, we will show only the results
of the maximal slicing condition hereafter.

The maximal slicing condition is known as a robust gauge condition
for singularity avoidance (or, exactly speaking,
avoiding the vanishing of the
volume elements of the associated Eulerian observers) \cite{york79}.
This is because the lapse
will go quite small value in the strong gravitational field.
Contrary, we may
guess whether the space-time will collapse or not
by monitoring the lapse function.

We plot the lapse function, $N$, in Fig.\ref{fig-lapse}.
Fig.\ref{fig-lapse}(a) is the lapse function
at the bubble surface,
$r=r_+$, versus time.
We see the lapse evolves small value for the
case of negative mass bubble space-time.
The lines end at the time when the violation of the constraint
equation begin growing.
{}From above standard behavior of the maximal sliced lapse functions,
we may say that the negative mass bubble space-time is `collapsing'
in some senses.
Fig. \ref{fig-lapse}(b) is snapshots of $N$ at several time
for the case of negative mass bubble space-time.

\subsection{Riemann invariant}
In order to confirm our guess of the `collapsing' behavior of the
negative mass bubble, we calculated the Kretchman invariant (Riemann
invariant) $R_{ijkl}R^{ijkl}$ of both 4 and 5-dimensional
 Riemann tensor.
   We see both blow up in the cases of negative mass
bubbles.
We plotted a typical behavior of the invariant
as a function of $r$ and $t$, in Fig.\ref{fig-rinv}.
These lines suggest that the possibility of the
formation of singularity in the final phase of evolution.

\subsection{Apparent horizons}
In order to confirm whether a black hole is formed or not in such a
case,
we check the appearance of
the apparent horizon.
The location of the apparent horizon is given by the position that
the expansion rate, of the outgoing null geodesic congruence
turns into the negative.
The appearance of the apparent horizon
indicates the existence of the event horizon.

The definition of the apparent horizon might not be unique in our
five-dimensional space-time
because it depends on the dimension of the space-time which the
null geodesic congruence runs\footnote{In usual
Kaluza-Klein picture, it is natural that the null geodesic congruence
runs in four dimensional part.}.
If the null congruence propagating in full five dimensional space-time,
the expansion rate, ${}^{\! (4)}\theta_+$, is given by
\begin{eqnarray}
{}^{\! (4)}\theta_+ &=& {}^{\! (4)}\nabla_a s^a
- {}^{\! (4)}K + s^a s^b  {~}^{\! (4)}K_{ab} \nonumber \\
&=& (a'+2c'+{2\over r}+{U' \over 2U}) \sqrt{U} e^{-b}
- (K^\chi_\chi+K^\theta_\theta+K^\varphi_\varphi)\label{deftheta}
\end{eqnarray}
where $s^a=(0,1/\sqrt{g_{rr}},0,0)$ is a outer pointing vector
in our spatial four
metric. 
On the other hand, if the null is confined to non-compactified four
dimensions, the expansion rate,
${}^{\! (3)}\theta_+$, is given similarly by
\begin{eqnarray}
{}^{\! (3)}\theta_+ &=& {}^{\! (3)}\nabla_a s^a
- {}^{\! (3)}K + s^a s^b  {~}^{\! (3)}K_{ab} \nonumber \\
&=& 2(c'+{1\over r}) \sqrt{U} e^{-b} -
(K^\theta_\theta+K^\varphi_\varphi).
\end{eqnarray}

We analyzed both ${}^{\! (3)}\theta_+$ and
${}^{\! (4)}\theta_+$ in our
evolving space-time.  Surprisingly,
in all cases (positive and negative mass
bubbles), both expansions remain positive definite
everywhere as we show an
example in Fig.\ref{fig-ah}.
These suggest us no-appearance of apparent horizons.

\section{Discussion} \label{Summary}

We numerically studied the dynamical evolution
of the Brill and Horowitz's initial data which can have the
negative energy.  As the zero energy Witten's bubble
space-time, we show that the `bubbles' with negative energy
will expand by mean of area  
upto the time we stop
the simulations.
At first glance this result supports Corley and Jacobson's
conjecture.  However, from the facts that the curvature invariant blows up,
and no appearance of the apparent
horizon, we suspect that a formation of a naked singularity as
the final fate of Kaluza-Klein negative energy bubble 
\footnote{
An anonymous referee of this article pointed out the similarity 
of the positive and negative mass bubble results.  
However, from the results we obtained, we believe that there are
qualitative differences between positive and negative mass cases
in their dynamical behaviors.  (We remark that our simulations 
are only up to a finite time in order to keep the resolution against
the expansion of the spacetime.)
}.
Hence, we may have to consider seriously the decay problem from
the Kaluza-Klein vacuum to the Witten-type `bubble' space-time.
Possible resolution to this may be given by assuming the
supersymmetry which may forbid the decay \cite{Witten,Marika},
or by constructing quantum gravity theory which may smooth
out singularities as normally been expected.

Although the negative mass bubbles are expanding, we obtained the
result that the bubble spacetime terminates at the singularity.
At first glance, they are incompatible, because the naive picture,
which the expanding keeps regularity. However, the picture may be
based on the Raychaudhri-type equation and the equation does not hold
in the present case. Moreover, the area 
cannot
properly describes whether the system will collapse or not.
Properly speaking, we need the proper radius from the center which
is absent in the present case.

{}Finally, we would like to comment on the so-called brane world scenario
\cite{brane,Tess,recent}. The brane world is motivated by the
reduction from the M-theory to the $E_8 \times E_8$ heterotic superstring
theory. This reduction drastically changes the picture of the
reduction \cite{brane} because
`matters' are confined to the ten dimensions and gravitons are
propagating in the eleven-dimensions. Here we call the timelike
hypersurface, where matters are confined, by `brane'.
A plausible history of the compactification are still
quite actively discussed recently and we do not reach the consensus
at this moment.
Apart from this compactification scenario, the reduction from ten to
four-dimensions follows the well-known Kaluza-Klein type or
Calabi-Yau compactification. Thus, our
present analysis is basically applicable to the space-time on the
brane, because we supposed the well-known Kaluza-Klein compactification.
More precisely, we may say that the  Witten-type Kaluza-Klein
`bubble' space-time on the 4-brane will be reduced from at least
6 dimensional space-time.
As was recently reported \cite{Tess},
the effective Einstein equations on the brane are different from the normal
Einstein equations. Therefore,
it might be worth re-asking what is the final fate of Kaluza-Klein
bubble if we describe the space-time by
 such a modified Einstein equation when we
take  a brane world scenario.

\section*{Acknowledgements}

TS is grateful to Gary Gibbons
and DAMTP relativity group for their hospitality.
HS appreciates  Abhay Ashtekar and Pablo Laguna for helpful comments,
and CGPG group for their hospitality.
HS also thank Ethan Honda for describing his numerical method.
Numerical computations were performed using machines at CGPG.
Both authors are supported by
the Japan Society for the Promotion of Science (JSPS) as research
fellows abroad.

\appendix
\section{The dynamical equations at the surface of the bubble}
In this appendix, we show
that $\dot{a}=\dot{b}$ at the location of the bubble,
$r=r_+$, during the time evolution.

The explicit expression of
Eqs (\ref{eq:einstein1}) and (\ref{eq:einstein2}) are given by
%
\begin{eqnarray}
\Bigl( \frac{\dot a}{N}\Bigr)^{\cdot}& =& -Ne^{-2b}
\Bigl[ \Bigl( -a'^2-a''-2a'c'-\frac{2a'}{r} +a'b'\Bigr)U
+\Bigl(-\frac{3}{2}a'+\frac{1}{2}b'-c'-\frac{1}{r}
 \Bigr)U'-\frac{1}{2}U''
\Bigr]\nonumber \\
& & -\frac{1}{N}({\dot a}+{\dot b}+2 {\dot c}){\dot a}
+e^{-2b}\Bigl(a'U+\frac{1}{2}U'\Bigr)N',
\\
\Bigl( \frac{\dot b}{N}\Bigr)^{\cdot}& =& -Ne^{-2b}
\Bigl[ \Bigl(
-a'^2-a''-2c'^2-2c''-\frac{4c'}{r}
+a'b'+2b'c'+\frac{2b'}{r}
\Bigr)U
+\Bigl(-\frac{3}{2}a'+\frac{1}{2}b'-c'-\frac{1}{r}  \Bigr)U'
-\frac{1}{2}U''
\Bigr]
\nonumber \\ & &
-\frac{1}{N}({\dot a}+{\dot b}+2 {\dot c}){\dot b}
+e^{-2b}\Bigl[(N''-b'N')U +\frac{1}{2}N'U' \Bigr],
\\
\Bigl( \frac{\dot c}{N}\Bigr)^{\cdot}& =& -Ne^{-2b}
\Bigl[ \Bigl(-2c'^2-c''-\frac{4c'}{r}-\frac{1}{r^2}+c'b'+\frac{b'}{r}
-a'c'-\frac{a'}{r} \Bigr)U
+\Bigl(-c'-\frac{1}{r} \Bigr)U'+\frac{e^{2b-2c}}{r^2}\Bigr]
\nonumber \\ & &
-\frac{1}{N}({\dot a}+{\dot b}+2 {\dot c}){\dot c}
+e^{-2b}\Bigl(c'+\frac{1}{r} \Bigr)N'U.
\end{eqnarray}
%
At $r=r_+$, we can truncate $U$, since $U=0$.  By adding a suffix $+$
for the variables which is evaluated at $r=r_+$,
the above equations become
%
\begin{eqnarray}
\Bigl( \frac{\dot a_+} {N_+} \Bigr)^{\cdot} & = & -N_+ e^{-2b_+}
\Bigl[ \Bigl(-\frac{3}{2}a_+'+\frac{1}{2}b_+'-c_+'-\frac{1}{r_+}
\Bigr)U_+'-\frac{1}{2}U_+'' \Bigr] \nonumber \\
& & -\frac{1}{N_+}({\dot a}_++{\dot b}_++2 {\dot c}_+){\dot a}_+
+\frac{1}{2}e^{-2b_+}U_+'N_+', \label{eq:aboundary}
\\
\Bigl( \frac{\dot b_+}{N_+}\Bigr)^{\cdot} & = & -N_+e^{-2b_+}
\Bigl[ \Bigl(-\frac{3}{2}a_+'+\frac{1}{2}b_+'-c_+'-\frac{1}{r_+}
\Bigr)U_+'-\frac{1}{2}U_+'' \Bigr] \nonumber \\
& & -\frac{1}{N_+}({\dot a}_++{\dot b}_++2 {\dot c}_+){\dot b}_+
+\frac{1}{2}e^{-2b_+}U_+'N_+'. \label{eq:bboundary}
\end{eqnarray}
%
Subtracting Eq. (\ref{eq:bboundary}) from Eq. (\ref{eq:aboundary}),
we obtain
%
\begin{eqnarray}
\Bigl( \frac{{\dot a}_+-{\dot b}_+}{N_+}\Bigr)^{\cdot}=-({\dot a}_+
+{\dot b}_++
2{\dot c}_+)\frac{{\dot a}_+-{\dot b}_+}{N_+}.
\end{eqnarray}
%
Therefore we get
%
\begin{eqnarray}
\frac{{\dot a}_+-{\dot b}_+}{N_+}=Ae^{-(a_++b_++2c_+)},
\end{eqnarray}
%
where $A$ is a constant.
Since the initial conditions,
(\ref{eq:initial1}) and  (\ref{eq:initial2}), imply
$ a_+=b_+$, which implies $A=0$. Therefore at the boundary, $r=r_+$,
we can set $ a_+=b_+$
even after the long time integration.


\newpage

{\bf Figure Captions}

\vspace{1.0cm}

\noindent
{\bf Fig.1}\\
Acceleration of the bubble surface, (\ref{accel}), versus time.
The figures (a) and (b) are of the geodesic slicing condition and
of the maximal slicing condition, respectively.  In both figures,
we see that the  negative mass bubble will soon be in collapse
phase, although they start expanding initially.
We set $q=1$ (hereafter for all figures). \\
For the case of evolutions with the maximum slicing condition, we plot
(c) area 
of the bubble surface $A$, (\ref{circumf}), versus time, and
(d) the velocity of the bubble surface, $dA/dt$.

\vspace{1.0cm}

\noindent
{\bf Fig.2}\\
The lapse function, $N$, are plotted
(the solutions of maximal slicing condition).
The figure (a) is $N$ at $r=r_+$ versus time.
We see the lapse evolves small value for the
case of negative mass bubble space-time.
The figure (b) is snapshots of $N$ at several times
for a case of negative mass bubble ($m=-0.4$) space-time.

\vspace{1.0cm}

\noindent
{\bf Fig.3}\\
(a) Typical snapshots of the Riemann invariant $R_{ijkl}R^{ijkl}$
of 4-dimensional
Riemann curvature for the case of negative mass bubble
($m=-0.4$) are plotted.
Only the region near the bubble surface is drawn.
(b) The
Riemann invariant $R_{ijkl}R^{ijkl}$ of 5-dimensional
Riemann curvature
are plotted as a function of time.
We see blow-ups in the cases of negative mass
bubbles (we cut the display range at $10^5$). 
The values are evaluated at a point right from
the bubble surface (that is, at $r_++\Delta r$).

\vspace{1.0cm}

\noindent
{\bf Fig.4}\\
A typical sample of the outgoing null expansion rate ${}^{\! (3)}\theta_+$
and ${}^{\!(4)}\theta_+$ are plotted for the case of negative mass
($m=-0.4$) bubbles.

\newpage
\begin{figure}[p]
\setlength{\unitlength}{1in}
\begin{picture}(7.0,7.0)
\put(1.5,4.25){\epsfxsize=4.0in \epsfysize=2.38in \epsffile{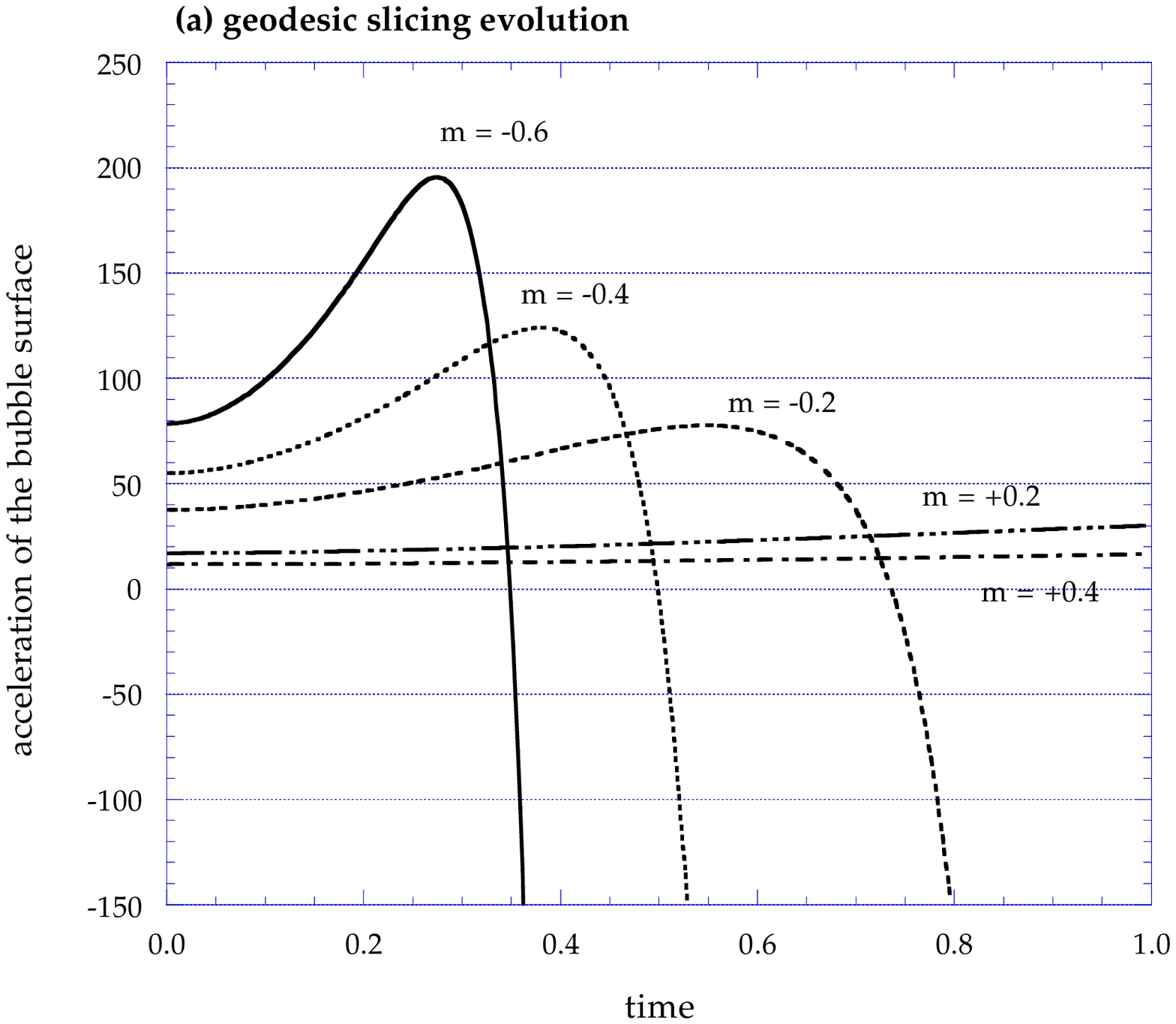} }
\put(1.5,0.25){\epsfxsize=4.0in \epsfysize=2.38in \epsffile{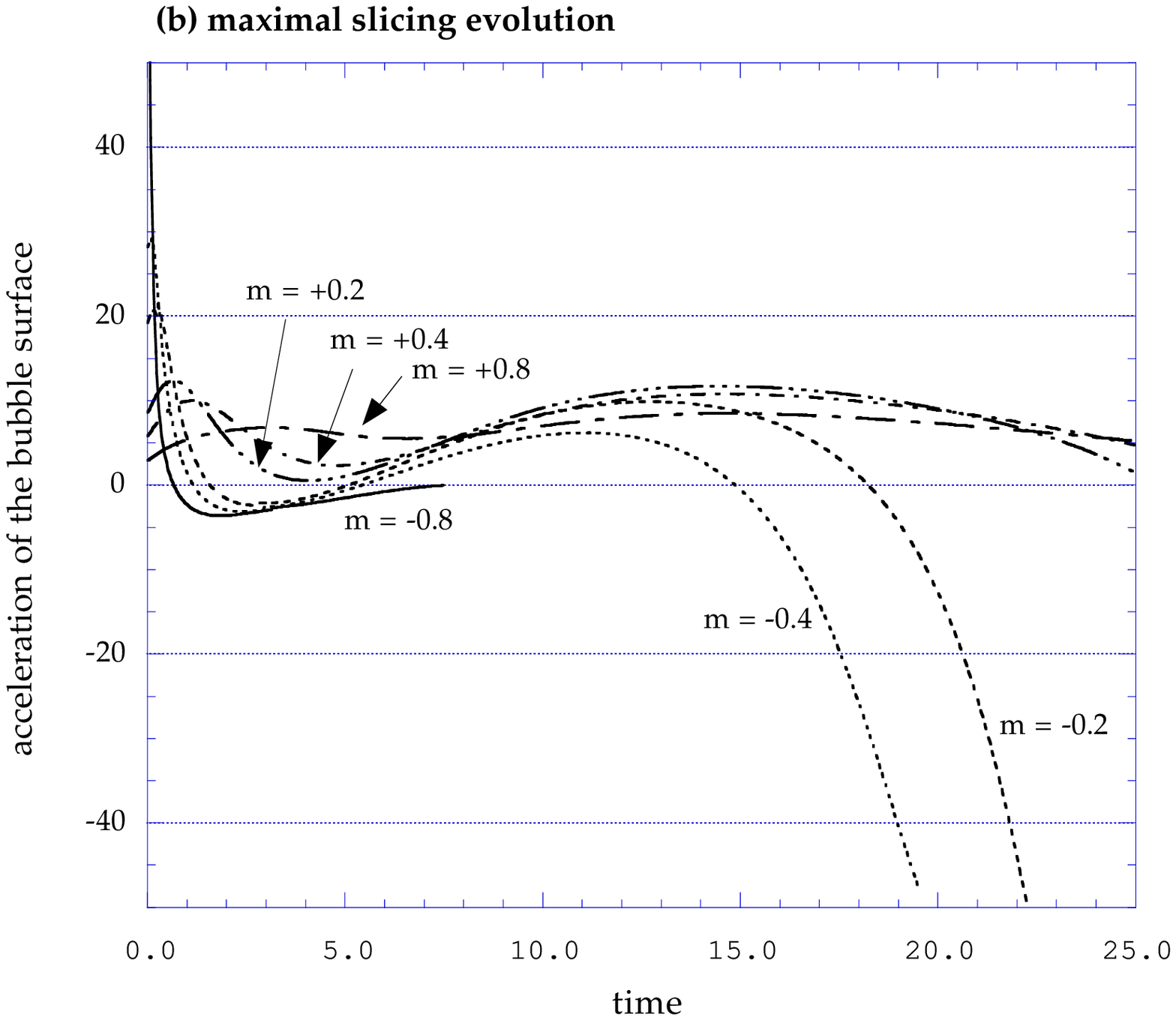} }
\end{picture}

\caption[fig-acc]{
Acceleration of the bubble surface, $d^2A/dt^2$, 
versus time.
The figures (a) and (b) are of the geodesic slicing condition and
of the maximal slicing condition, respectively.  In both figures,
we see that the  negative mass bubble will soon be in collapse
phase, although they start expanding initially.
We set $q=1$ (hereafter for all figures).
}
\label{fig-acc}
 \end{figure}
\newpage
\setcounter{figure}{0}
 \begin{figure}[p]
\setlength{\unitlength}{1in}
\begin{picture}(7.0,7.0)
\put(1.5,4.25){\epsfxsize=4.0in \epsfysize=2.38in \epsffile{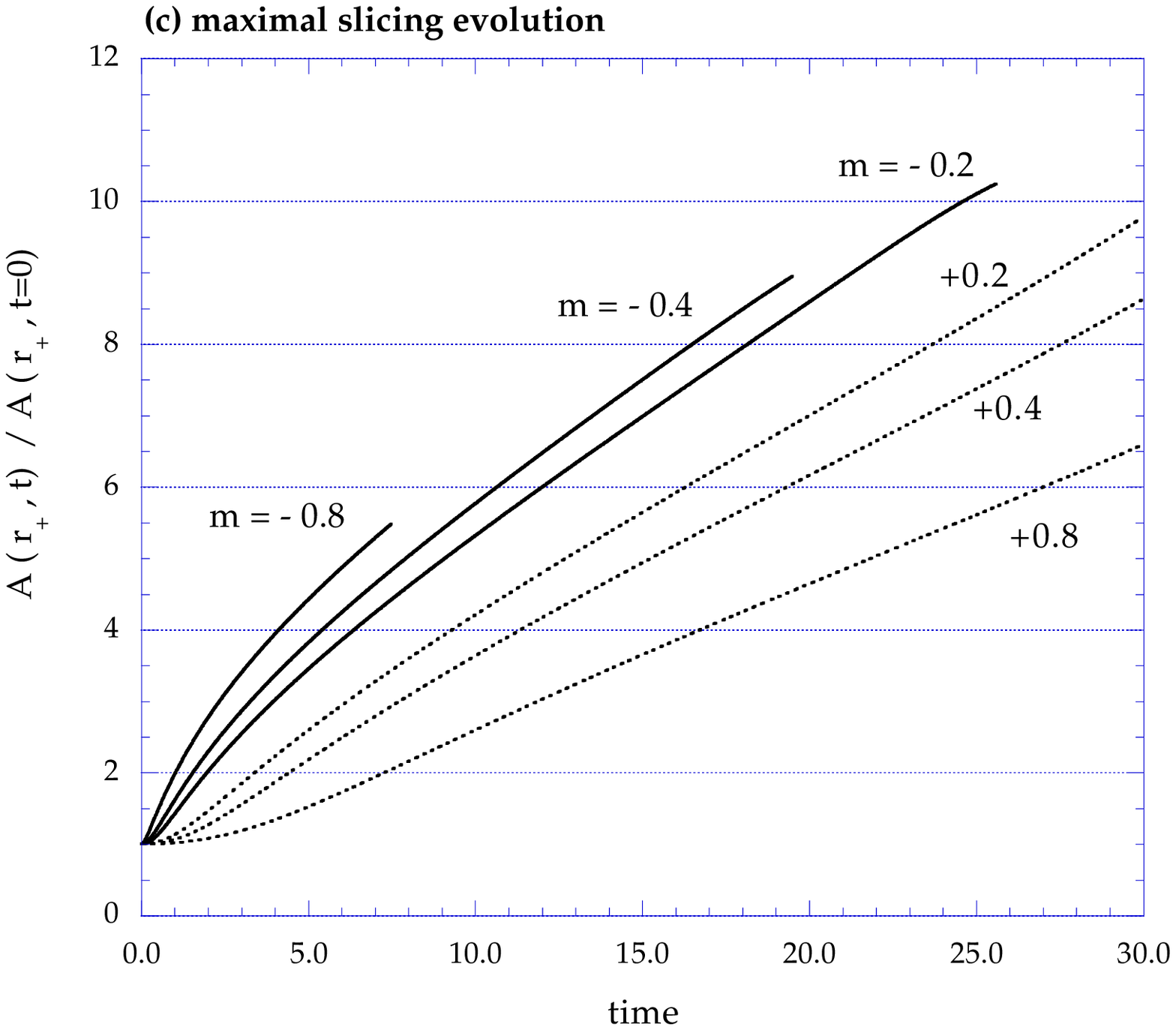} }
\put(1.5,0.25){\epsfxsize=4.0in \epsfysize=2.38in \epsffile{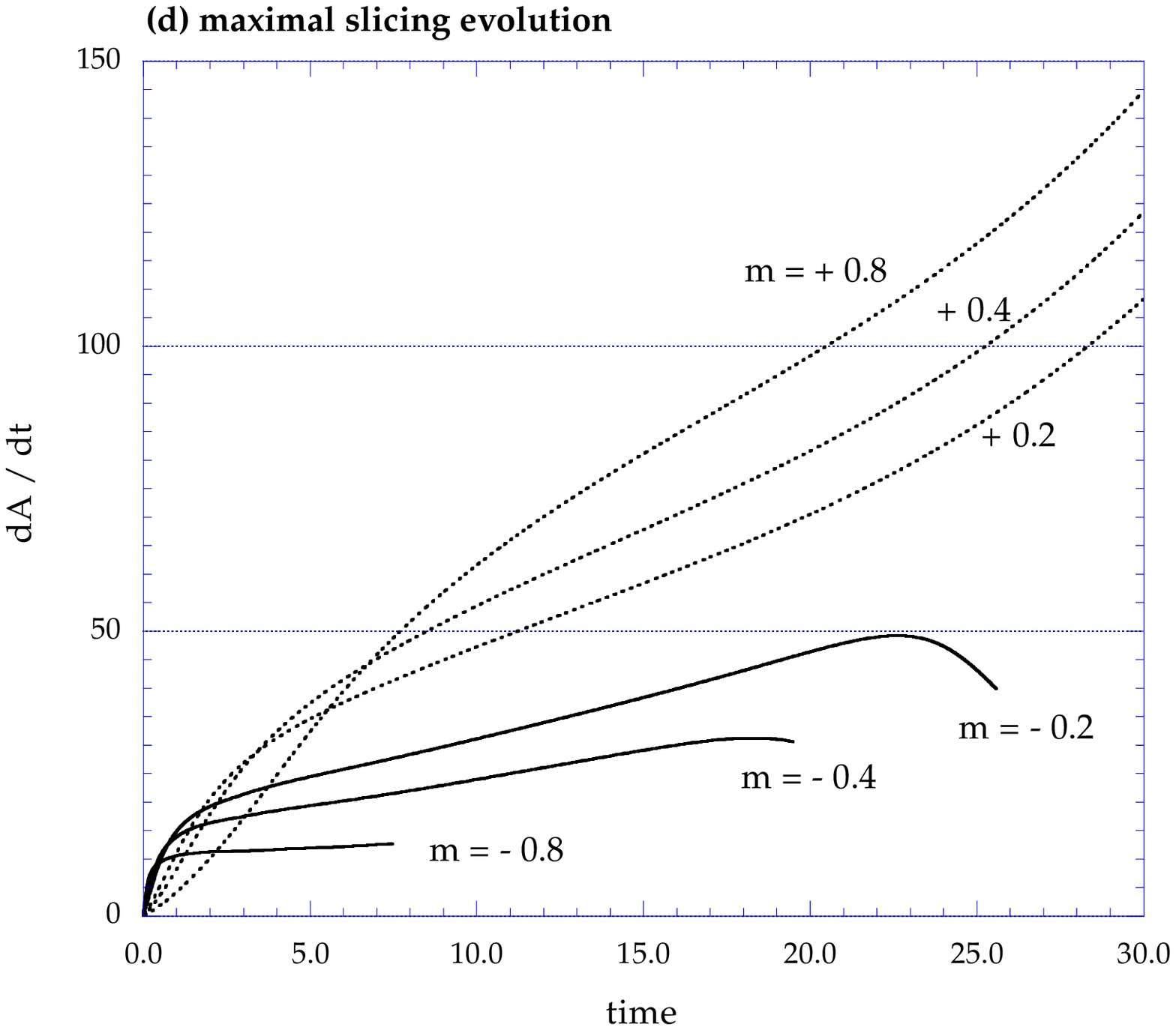} }
\end{picture}

\caption[fig-acc]{
(continued)\\
For the case of evolutions with the maximal slicing condition, we plot
(c) area 
of the bubble surface $A$, (\ref{circumf}), versus time, and
(d) the velocity of the bubble surface, $dA/dt$.
}
\end{figure}
\newpage

\begin{figure}[p]
\setlength{\unitlength}{1in}
\begin{picture}(7.0,7.0)
\put(1.5,4.25){\epsfxsize=4.0in \epsfysize=2.38in \epsffile{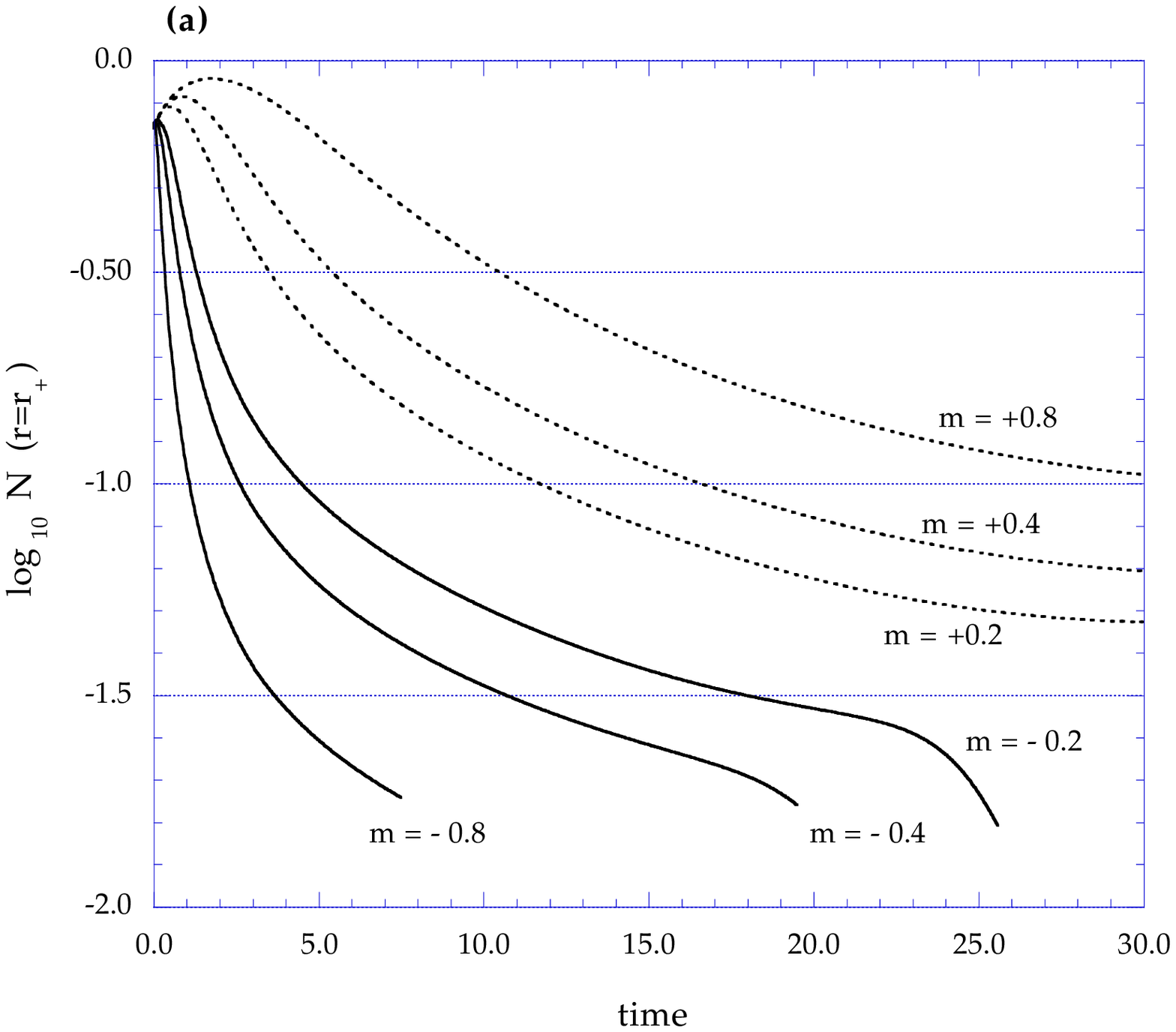} }
\put(1.5,0.25){\epsfxsize=4.0in \epsfysize=2.38in \epsffile{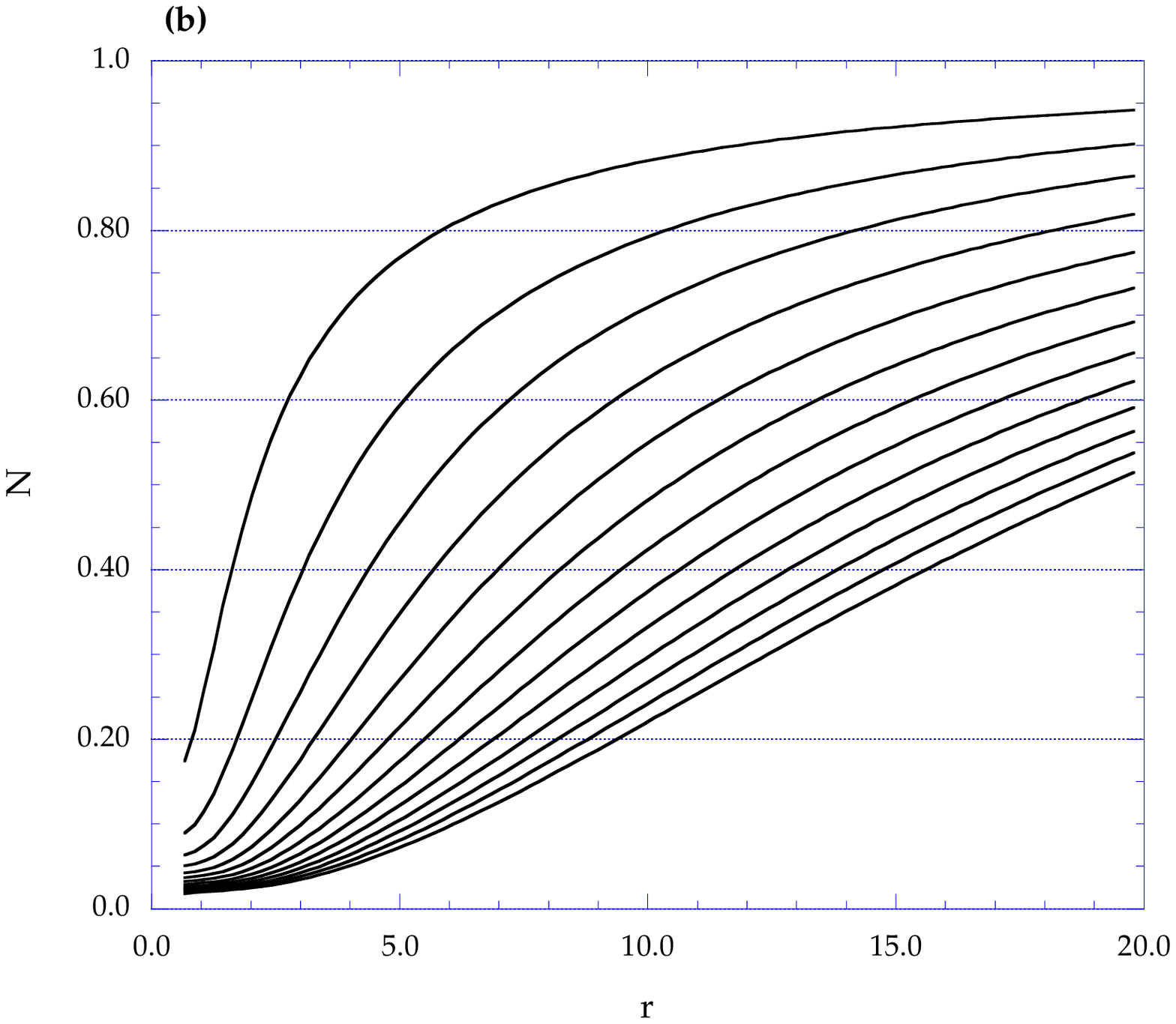} }
\end{picture}

\caption[fig-lapse]{
The lapse function, $N$, are plotted
(the solutions of maximal slicing condition).
The figure (a) is $N$ at $r=r_+$ versus time.
We see the lapse evolves small value for the
case of negative mass bubble space-time.
The figure (b) is snapshots of $N$ at several times
for a case of negative mass bubble ($m=-0.4$) space-time.
}
\label{fig-lapse}
\end{figure}

\begin{figure}[p]
\setlength{\unitlength}{1in}
\begin{picture}(7.0,7.0)
\put(1.5,4.25){\epsfxsize=4.0in \epsfysize=2.38in \epsffile{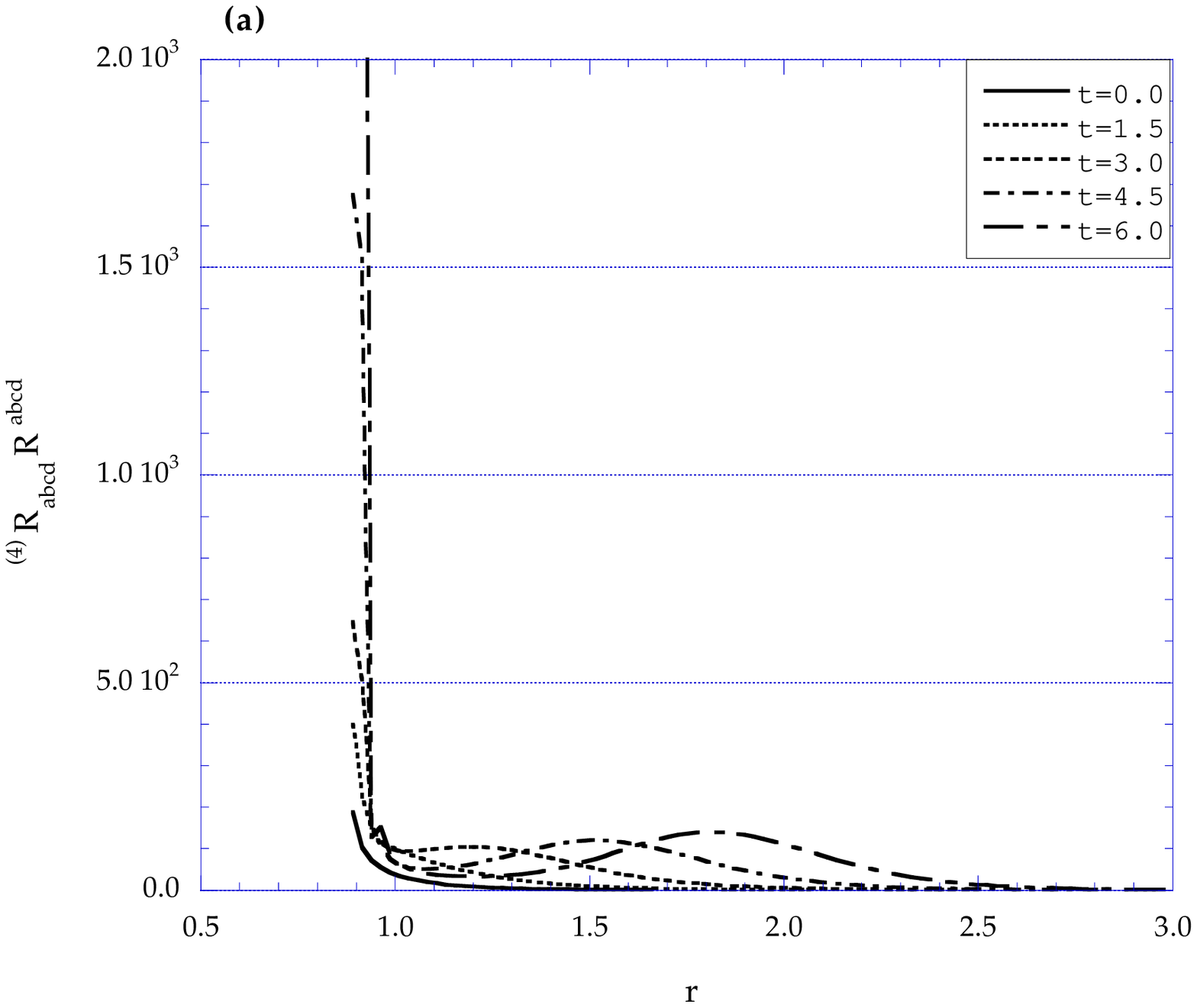} }
\put(1.5,0.25){\epsfxsize=4.0in \epsfysize=2.38in \epsffile{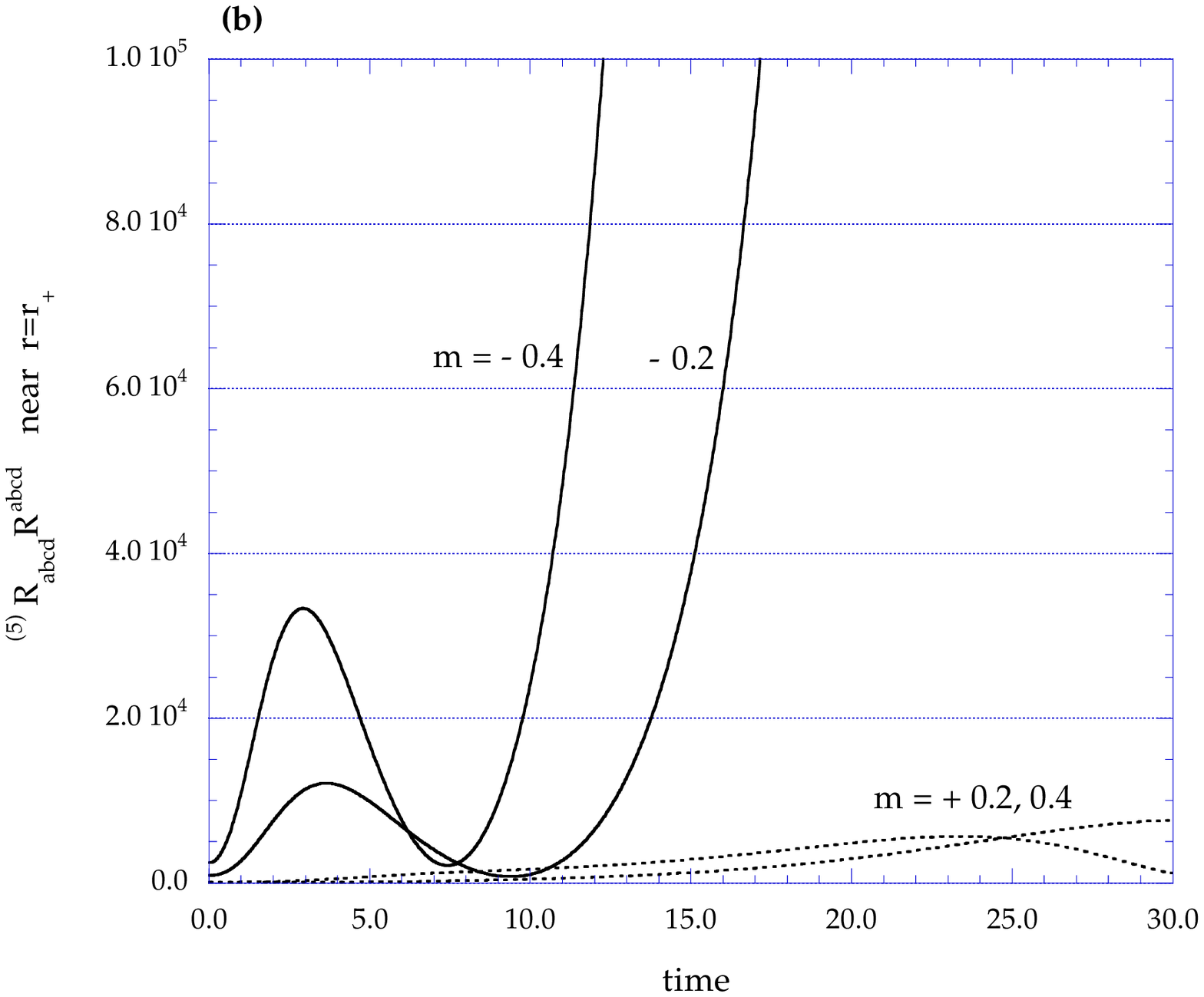} }
\end{picture}

\caption[fig-rinv]{
(a) Typical snapshots of the Riemann invariant $R_{ijkl}R^{ijkl}$
of 4-dimensional
Riemann curvature for the case of negative mass bubble
($m=-0.4$) are plotted.
Only the region near the bubble surface is drawn.
(b) The
Riemann invariant $R_{ijkl}R^{ijkl}$ of 5-dimensional
Riemann curvature
are plotted as a function of time.
We see blow-ups in the cases of negative mass
bubbles (we cut the display range at $10^5$). 
The values are evaluated at a point
right from the bubble surface (that is, at $r_++\Delta r$).

}
\label{fig-rinv}
\end{figure}

\begin{figure}[p]
\setlength{\unitlength}{1in}
\begin{picture}(7.0,7.0)
\put(1.5,4.25){\epsfxsize=4.0in \epsfysize=2.38in \epsffile{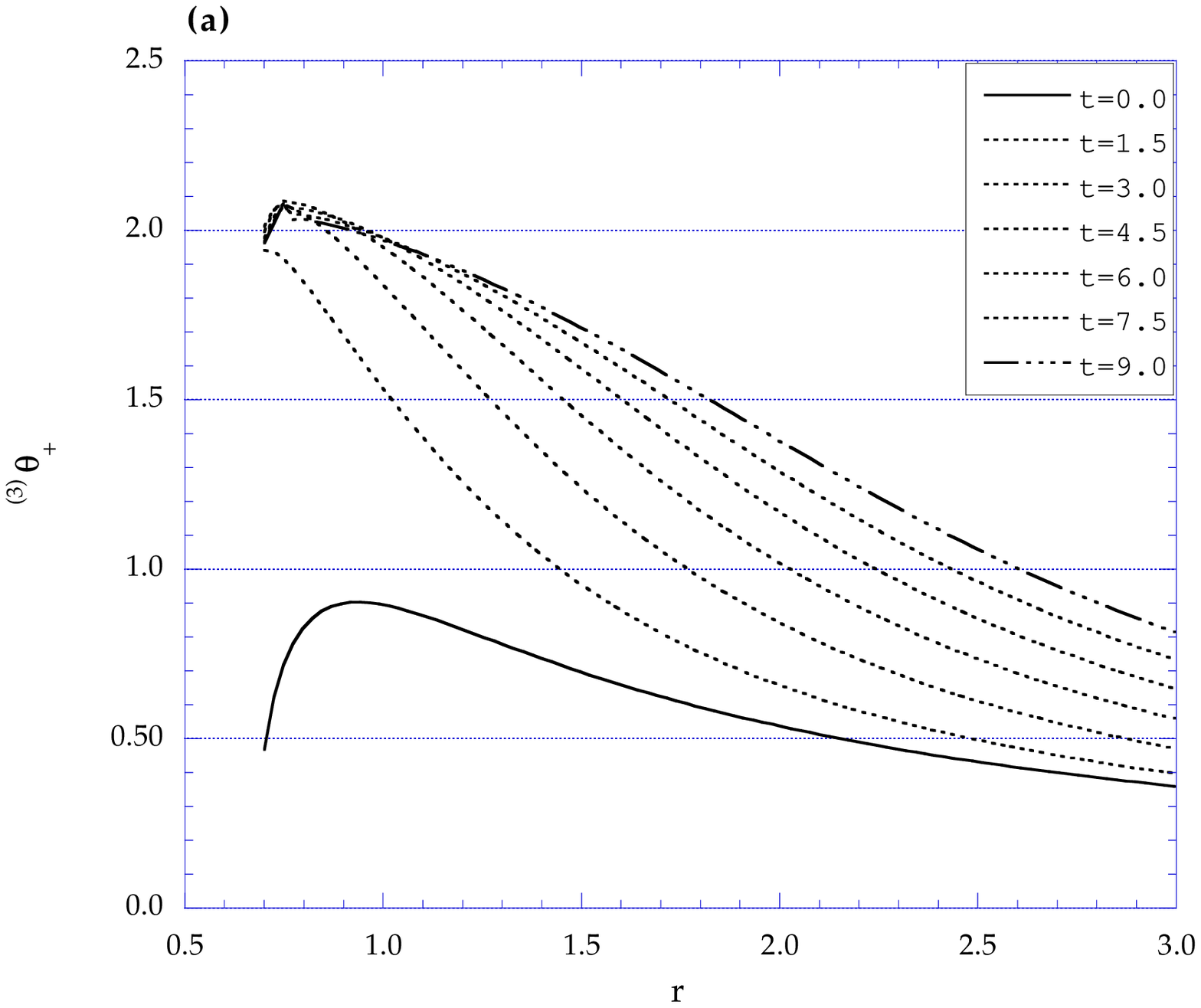} }
\put(1.5,0.25){\epsfxsize=4.0in \epsfysize=2.38in \epsffile{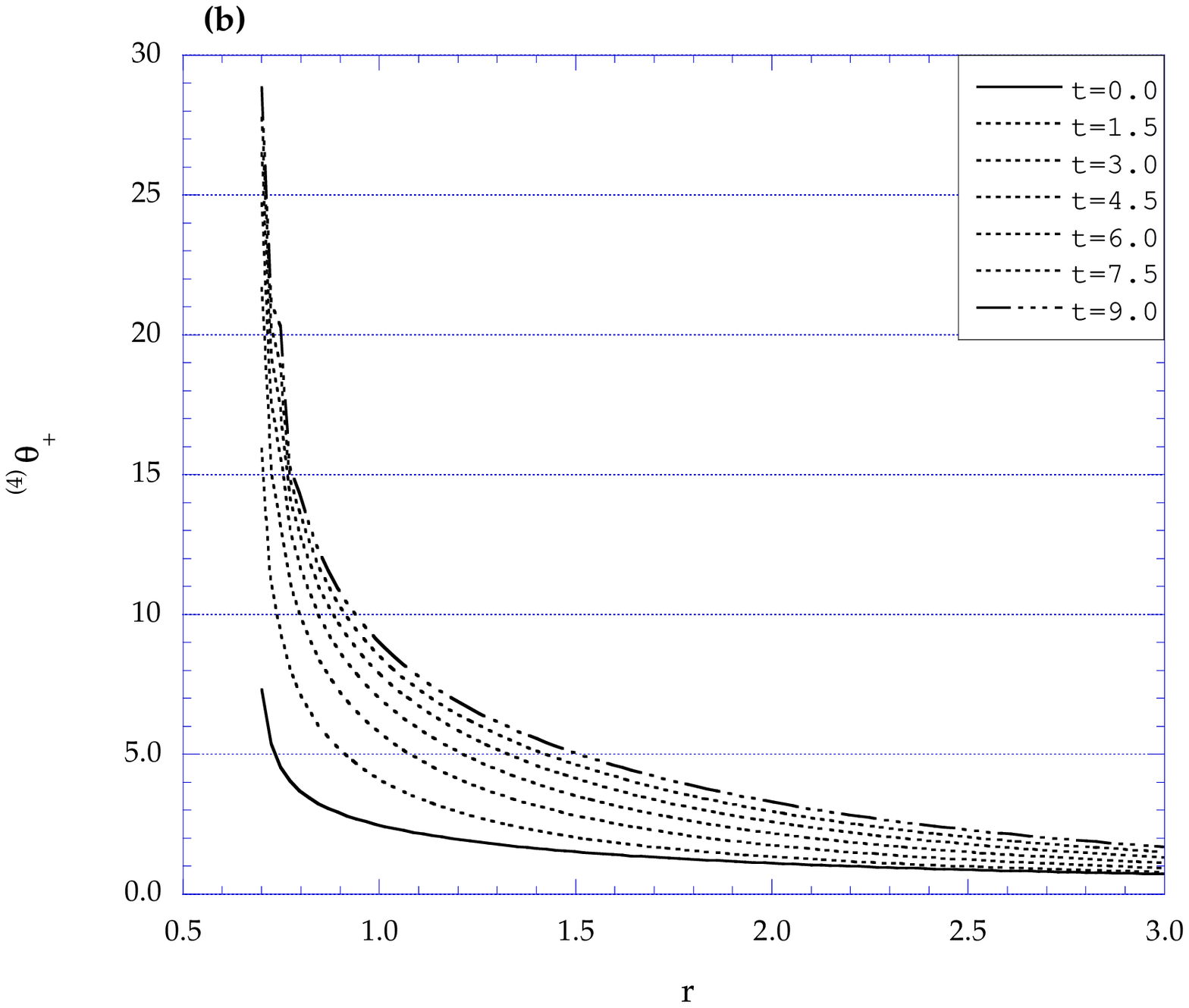} }
\end{picture}

\caption[fig-ah]{
A typical sample of the outgoing null expansion rate ${}^{\! (3)}\theta_+$
and ${}^{\!(4)}\theta_+$ are plotted for the case of negative mass
($m=-0.4$) bubbles.
}
\label{fig-ah}
\end{figure}

\end{document}